\documentclass{ws-ijmpb}

\begin{document}

\markboth{J. R. Soto, J. J. Castro, E. Y\'epez, and A. Calles} {DF
electronic spectrum of the CuO${_{6}}^{-10}$ and JT distortions in La-Ba-Cu-O.}

%
\catchline{}{}{}{}{}
%

\title{DENSITY FUNCTIONAL ELECTRONIC SPECTRUM OF THE CuO${_{6}}^{-10}$
CLUSTER AND POSSIBLE LOCAL JAHN-TELLER DISTORTIONS IN THE La-Ba-Cu-O SUPERCONDUCTOR.}

\author{J. R. SOTO$^1$, J. J. CASTRO$^2$, E. Y\'EPEZ$^1$, AND A. CALLES$^1$\footnote{ Corresponding author:
calles@servidor.unam.mx}}

\address{$1$ Facultad de Ciencias, Universidad Nacional Aut\'onoma de M\'exico,\break
Apdo. Post. 70-646, M\'exico D. F., 04510, M\'exico. \\ $2$ Departamento de F\'{\i}sica,
Centro de Investigaci\'on y Estudios Avanzados del IPN, \break Apdo. Post. 14-740,
M\'exico D. F., 07000 , M\'exico.}

\maketitle

\begin{history}
\received{Day Month Year} \revised{Day Month Year}
\end{history}

\begin{abstract}

We present a density functional theory (DFT) calculation in the generalized gradient
approximation to study the possibility for the existence of Jahn-Teller (JT) or pseudo
Jahn-Teller (PJT) type local distortions in the La-Ba-Cu-O superconducting system. We
performed the calculation and correspondingly group theory classification of the
electronic ground state of the CuO${_{6}}^{-10}$ elongated octahedra cluster, immersed
in a background simulating the superconductor. Part of the motivation to do this study
is that the origin of the apical deformation of the CuO${_{6}}^{-10}$ cluster is not due
to a pure JT effect, having therefore a non {\it a priori} condition to remove the
degeneracy of the electronic ground state of the parent regular octahedron. We present a
comparative analysis of the symmetry classified electron spectrum with previously
reported results using unrestricted Hartree-Fock calculations (UHF). Both the DFT and
UHF calculations produced a non degenerate electronic ground state, not having therefore
the necessary condition for a pure JT effect. However, the appearance of a degenerate
E$_{g}$ state near to the highest occupied molecular orbital in the DFT calculation,
suggests the possibility for a PJT effect responsible for a local distortion of the
oxidized CuO$_{6}^{-9}$ cluster.

\end{abstract}


\section{Introduction}

\label{intro}

Since the discovery of high T$_{c}$ superconductivity, the understanding of the
mechanism responsible for this novel phenomenon has been one of the most challenging
problems in condensed matter physics. To date, it is known to involve the pairing of
charge carriers, although the precise nature of these carriers and the mechanism by
which the pairing occurs is still an open problem.\cite{Legg}$^{-}$\cite{Zha} The
evidence that strong electron-phonon interaction in oxides as well as in mixed-valence
systems can occur owing to polaron formation, together with the fact that a possible
mechanism for polaron formation is the Jahn-Teller (JT) effect,\cite{JT} where
spontaneous lattice distortions remove the degeneracy of the electronic ground state,
led to Bednorz and M\"{u}ller\cite{BM} to the discovery of high temperature
superconductivity. One reason to suspect the JT effect might play an important role in
those systems, is the fact that all superconducting copper oxides have perovskite-like
crystal lattices, with Cu$^{2+}$ ions surrounded by O$^{2-}$ ions. One of the most
studied system, La-Ba-Cu-O has this particular structure, containing sheets of
corner-sharing CuO$_{6}$ octahedra. The Cu-O planes, where it is believed that
superconductivity and charge transport are mostly confined, have an electronic
structure, within the ionic model, built up from the 3d$^{9}$ states of Cu$^{2+}$ and
the 2p$^{6}$ states of O$^{2-}$.

In the insulating parent compound of the La based superconductor La$_{2}$CuO$%
_{4}$, the CuO${_{6}}$ octahedra are elongated, with a Cu-O distance of 1.90~\AA
\hspace{0.1cm} within the plane and 2.42~\AA \hspace{0.1cm} perpendicular to it. This
distortion, some authors believed, could be at least partly due to a static JT
effect.\cite{OB} This can be seen from the fact that for the regular octahedron we have
the O$_{h}$ symmetry, with a splitting of the d$^{9}$ electron states into E$_{g} +$
T$_{2g}$ irreducible representation (IR). The corresponding decomposition of the ions
vibrational modes $Q_{k}$, according to the IR of the O$_{h}$ group gives $\Gamma =$
a$_{1g}+$ e$_{g}+$ t$_{2g}+2$t$_{1u}+$ t$_{2u}$. The general condition in order to have
a JT effect, is that the generalized
force between nuclei, defined as the matrix element of the force operator (-$%
\partial V/\partial Q_{k}$), between the total electronic ground
state $\psi _{i}$, transforming respectively as the $\Gamma _{k}$ and $\Gamma _{i}$ IR
of the corresponding symmetry group, should be different from zero. From general group
theory rules, we know that this condition can be fulfilled if $\Gamma _{i}^{*}\times
\Gamma _{k}\times \Gamma _{i}\subset \Gamma _{1}$ is satisfied (or equivalently $\Gamma
_{i}^{*}\times \Gamma _{i}\subset \Gamma _{k}$), where $\Gamma _{1}$ is the identity IR.
Hence for the regular octahedron having the O$_{h}$ symmetry group, the JT condition is
fulfill, being the e$_{g}$ and t$_{2g}$ modes the possible JT active modes. Hence in
order to have a pure JT effect and as a consequence a completely removal of the
electronic degeneracy, the Cu site symmetry should be cubic. This would be precisely the
case for an isolated regular CuO$_{6}$ octahedra. However
in the insulating parent compound of the La based superconductor La$_{2}$CuO$%
_{4}$, the Cu site symmetry is far from being cubic. Even if the CuO$%
_{6} $ octahedra would be regular, the crystal field would not possess cubic symmetry.
It has also been shown\cite{Pic} (and references cited there), that at least half of the
octahedral distortion in the CuO$_{6}$ cluster can be accounted for, as a consequence of
the strongly layered crystal structure and the large ionic interaction. From the above
mentioned arguments one would expect that the partially filled electronic E$_{g}$ state
in the regular CuO$_{6}$ octahedra might only be partly responsible for the CuO$_{6}$
elongated shape, having therefore a non {\it a priori} condition for the complete
removal of the electronic degeneracy. As a clear example of an apical distortion where
the JT effect does not play any role is when the Cu ion is substituted by Ni, obtaining
La$_{2}$NiO$_{4}$. This system has the same structure as La$_{2}$CuO$_{4}$, but the
Ni$^{+2}$ is not a JT ion. There are also in the literature some proposed models where
the JT effect plays a definite role in high T$_{c}$ superconductivity beyond that of
producing the static deformation of the CuO$_{6}$ octahedra.\cite{Eng}$^{-}$\cite{Tsch}

Experimental results in Cu K-edge absorption spectroscopy XANES (x-ray absorption near
edge structure) and EXAFS\cite{Saini} (extended x-ray absorption fine structure) as well
as PDF (pair-distribution-function) analysis of neutron powder-diffraction
data\cite{Bazel} have shown the existence of important inhomogeneities in the high-Tc
superconductor structures. In particular, the CuO$_{6}$ octahedra in
La$_{1.85}$Sr$_{0.15}$CuO$_{4} $ exhibits local deformations in the apical direction as
well as in the planar Cu-O bond length due presumably to changes in the local charge
state.\cite{Saini} This fact together with the previously mentioned arguments concerning
the possible symmetry of the electronic ground state, raises the question as whether
this deformation is somewhere connected to a JT or pseudo Jahn-Teller (PJT) kind of
distortion. In order to determine whether the JT or PJT effect might play any
significant role on the local distortion on the La-Ba-Cu-O system, it is necessary to
determine the symmetry properties of the total electronic   ground state. Being the
condition for a JT that this state must be degenerate. If it is not the case then the
possibility of a PJT effect emerges and it is determined by
no null matrix elements of the force operator (-$%
\partial V/\partial Q_{k}$), between the non degenerate ground state
and closed excited  states. Concerning the PJT effect it has been pointed
out\cite{Bersuker} that there are no {\it a priori} exceptions for vibronic coupling
effects. The presence of electronic degeneracy is just an important special case of
vibronic coupling. Bersuker\cite{Bersuker2} has introduced a  PJT vibronic coupling
through the mixing with the $^{1}$B$_{1g}$ excited state to explain a polaron formation
mechanism in high T$_{c}$ superconductors. It is worth mention that the PJT effect has
been some times misinterpreted as a second order correction, and in fact this effect may
be very strong, specially when the JT effect is absent. The study of those effects might
be relevant in understanding the electron-lattice interaction.

In this paper, we examine the possibility of having a local distortion in the CuO$_{6}$
octahedra in La-Ba-Cu-O superconducting system originated by a JT or PJT effect. This is
done under the assumption of a CuO${_{6}}^{-10}$ cluster model for the superconductor.
The CuO${_{6}}^{-10}$ cluster is immersed in a set of static charges, located in the
proper positions in order to simulate the crystalline background. This background
stabilizes the cluster and preserves the point symmetry group. The electronic spectrum
is calculated using the density functional theory in the generalized gradient
approximation (DFT-GGA), and classified according to the IR of the symmetry group
D$_{4h}$.

The paper is organized as follows. In Sec. 2 we introduce the cluster model representing
the La-Ba-Cu-O system, and the methods used for the calculation and symmetry
classification of the electronic spectrum. In Sec. 3 we present the results and
discussion for the spectrum obtained and compared with the results of a unrestricted
Hartree-Fock (UHF) calculation, and discussed the possible JT or PJT type of
distortions. The conclusions are presented in Sec. 4.

\section{Cluster Model}

\label{cluster}

Embedded cluster  models are used to describe real systems by treating the cluster
electronic structure with some degree of sophistication by using {\it ab initio} or
semi-empirical methods, where as cruder approximations have to be made in order to treat
the background used to embed the cluster. The basic philosophy in the use of cluster
models is to study essential characteristics of the solid which are primarily determined
by local properties that can be simulated by cluster calculations. Its use has been
mainly justified by many successful applications including, by example, the
La$_{2}$CuO$_{4}$ undoped cuprate\cite{Renold,deGraaf} or the
La$_{2-x}$Sr$_{x}$CuO$_{4}$ superconductor system.\cite{Calzado} In this work we used
the cluster model to study the effect of the localized orbital electronic states on the
possibility for the existence of Jahn-Teller (JT) or pseudo Jahn-Teller (PJT) type local
distortions in the La-Ba-Cu-O superconducting system.

 We study the elongated octahedral
CuO${_{6}}^{-10}$ cluster immersed in a point charges environment according to the
structure of the tetragonal superconductor (La,Ba)$_{2}$CuO$_{4}$ with the experimental
lattice constants: $a=3.7873$~\AA \hspace{0.1cm} and $c=13.2883$~\AA \hspace{0.1cm} and
with a Cu-O apex distance of 2.42~\AA.\cite{Jorgensen} This background simulates the
crystalline surrounding.

The necessity to preserve the experimental bond lengths of the cluster, lead us to
adjust the background arrangement in order to stabilize the cluster system. The values
for the point charges environment were determined by performing a UHF cluster stability
analysis.\cite{SCCY} The first trial of stability calculations was done surrounding the
cluster with fixed point charges corresponding to the formal charges values La$^{3+}$,
Cu$^{2+}$, O$^{2-}$, and placing O$^{1.1538-}$ at the outer boundary (in a spherical
radius of 8~\AA \hspace{0.1cm}) accumulating 178 charges which neutralize the -10 charge
of the cluster. We found this structure unstable in this background. However, if we
change the formal charge of the nearest copper atoms in Cu-O planes to Cu$^{1+}$ (value
consistent with possible Cu valence states),
 we recuperate stability with an absolute value for the
energy gradient of 2.2 $\times $10$^{-3}$ a.u. for the oxygen atoms in the plane. At
this stage we did not try to make a full structure optimized calculation.

Recent works dealing with similar cuprate clusters have shown that the use of
pseudopotential models for the nearest atoms to the cluster are advantageous in order to
prevent an artificial polarization of the oxygen electrons towards the (in other way)
bare positive charges.\cite{deGraaf,Calzado,Husser} In a previously reported
work,\cite{SCCY} we performed an electronic structure calculation for the neutral
CuO${_{6}}^{-10}$ cluster embedded in the modified point charges environment using
different basis in the HF formalism. Looking at the hybridization of the different wave
functions, we do not observe the artificial polarization effect. This is due to the fact
that we are using an effective charge of +1 for the nearest Cu atoms on the Cu-O planes
. The only case where we note anomalies was in the use of minimal gaussian basis 3-21G,
where we got a wrong symmetry for the total wave function. By the addition of a
polarization term to the oxygen 3-21G basis, we recuperated the correct wave function
symmetry. It should be mentioned that the point charges substitution by the
pseudopotentials does not guarantee the stabilization of the structure without a full
relaxation of the cluster. In a DFT structure optimization calculation using bare
pseudopotentials, relaxing only the apical length, Pliber\v{s}ek et al.\cite{Meier}
found a +0.04~\AA \hspace{0.1cm} shift from the correct experimental value for the
apical distance Cu-O. Hence, since our main  interest is in the analysis of the symmetry
properties of the electron spectrum we try as a first model calculation a background
with point charges.

When a point charges set is used to simulate the crystalline environment, some small
adjustments in the charge values or its positions, or the inclusion of a ghost charges
set are made in order to obtain the Madelung potential on the points corresponding to
the atoms positions in the cluster. We do not assume this procedure here because the
difference between the potential resultant from the number of charges in our model and
the Madelung potential becomes rather small.\cite{Martin} Even more, assuming this
procedure, the background symmetry could be changed if some charges are moved from its
original positions. In our model we need to preserve the point symmetry of the
environment in order to obtain the correct IR\'{}s corresponding to the cluster
molecular orbital (MO) and the nature of the vibronic coupling.

In summary, our model consists of a CuO${_{6}}^{-10}$ cluster embedded in an environment
formed by point charges placed at the positions of the nearest shell of atoms outside
the cluster with the outer shell always being oxygen and the immediate copper atoms in
the Cu-O plane with a charge +1. For the present calculation we have surrounded the
cluster with 178 fixed point
charges corresponding to La$^{3+}$, Cu$^{2+}$, O$^{2-}$, Cu$^{1+}$, and O$%
^{1.026-}$ at the outer boundary, with a net charge of +10 which neutralized the -10
charge of the cluster. The radius of the outer shell was fixed at 8~\AA \hspace{0.1cm}.
In order to consider the possible undesirable basis effects on the electronic spectrum
we selected a 4-21G basis with polarization d-components in the oxygens and no diffuse
components s or p were added. The additional oxygen d-components privileges
the Cu3d-O2p hybridization over the artificial polarization O2p-(Cu$^{1+}$, La%
$^{3+}$ bare charges). The arrangement selected, guarantees approximately the stability
of the cluster, as well as the D$_{4h}$ symmetry properties of the electronic states and
small polarization effects.

 In a simple ionic model, the molecular levels formed by the 3d$%
^{9}$ states of Cu$^{2+}$ hybridized with the 2p$^{6}$ states of O$^{2-}$ immersed in a
D$_{4h}$ crystalline field produce a nondegenerate electronic ground state. The same
result is obtained within a quasi-molecular approximation in an extended H\"{u}ckel
model and also in open shell restricted and unrestricted Hartree-Fock (RHF-OS and UHF)
calculations for the electronic ground state.\cite{SCCY} This shows that within these
approximations we do not find a possibility for pure JT distortions. The question is
whether the same symmetry for the total electronic wave function is preserved in an
electronic spectra obtained from {\it ab initio} calculations where higher electron
correlations are taken into account. In order to investigate this situation we have
performed a DFT-GGA calculation to determine the electronic ground state  for the apical
octahedron cluster together with their symmetry classification. This allows us to
analyze the possibility of realization of total degenerate electronic wave function.

The DFT calculation for the electronic ground state was performed using {\it The
Cambridge Analytic Derivatives Package} CADPAC 6.1\cite{cad} running in a CRAY/YMP-4.
This calculation was done using the generalized gradient approximation
(GGA)\cite{Lang}$^{-}$\cite{Perde} with a 3-21G basis set for Cu and a 4-31G$^{*}$ for O
(where $^{*}$ means polarized). The GGA improves the previous commonly used DFT-LDA
(local density approximation) incorporating some nonlocal effects to the
exchange-correlation functional using the gradient of the electron density. We employed
a convergence criterium of 1$\times $10$^{-7}$ for the density. The expected value for
the spin operator S$^{2}$ in the ground state, was systematically checked to be 3/4. The
symmetry of the total electronic wave function and
hence of the density is guaranteed by the symmetry of the hamiltonian (D$%
_{4h}$). For the exchange-correlation X-C functionals, the BP86
approximation (Becke\cite{Becke} exchange and Perdew correlation\cite{Perd}%
) was used. We consider a system formed by 87 electrons of the CuO${_{6}}^{-10}$
octahedron cluster which corresponds to the Cu$^{2+}$ and 6O$^{2-}$ ions. For the
calculation, a total of 119 Gaussian base functions were
employed, corresponding to 29 for the Cu$^{2+}$ and 15 for each of the six O$%
^{2-}$.The number of $\alpha $ and $\beta $ occupied states are 44 and 43 respectively,
corresponding to the 87 electrons within the cluster. The obtained electronic wave
functions were classified according to the IR of the D$_{4h}$ symmetry group using a
simple algorithm based on the projection of characters that permits to find the
classification according to the IR of a given symmetry group with the application of at
most three symmetry operation.\cite{YCC}

\section{Results and Discussion}

\label{results}

\begin{figure}[th]
\centerline{\psfig{file=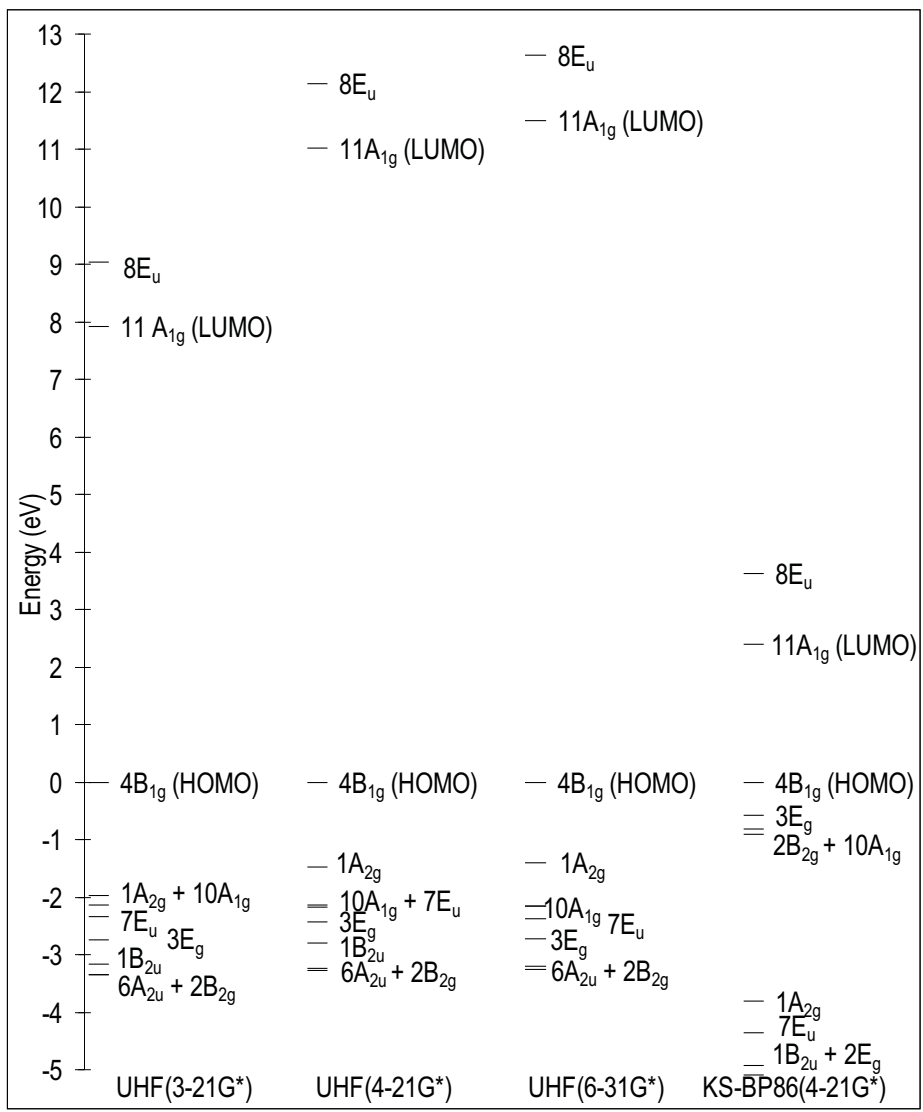,width=1.\textwidth}} \vspace*{8pt} \caption{DFT
electronic spectrum for the $\alpha$ states using the Cu-3-21G plus O-4-21G$^{*}$ basis
set. This is compared with the UHF spectra using the Cu-3-21G plus O-3-21G$^{*}$,
O-4-21G$^{*}$ and O-6-31G$^{*}$ basis sets respectively of the CuO${_{6}}^{-10}$
cluster.}
\label{F1}       
\end{figure}

We have studied within density functional theory in the generalized gradient
approximation, the electronic ground state for the CuO${_{6}}^{-10}$ elongated octahedra
cluster. This cluster represent one of the T$_{c}$ superconductors that have been
suggested as a prominent candidate to show the JT effect : La-Ba-Cu-O.

The electronic spectrum for $\alpha$ states in the DFT for the Cu-3-21G plus
O-4-21G$^{*}$ basis set in joint to the UHF spectra for Cu-3-21G plus O-3-21G$^{*}$,
O-4-21G$^{*}$ and O-6-31G$^{*}$ is shown in Fig.~\ref{F1}. The HOMO (highest occupied
molecular orbital) states correspond, in both UHF and DFT, to a nondegenerate $\alpha$
level whose IR is 4$B_{1g}$ giving, after making the product with the $\beta$ electronic
wave function, a total orbital nondegenerate electronic ground state wave function whose
symmetry is $^{2}$$B_{1g}$. It should be mentioned that in Ref.~\refcite{Husser},
although the authors do not explicitly refer to the symmetry of the ground state, it is
possible to infer it from their components of the atomic orbitals, corresponding the
$^{2}$B$_{1g}$, which agrees with our result. As can be seen from the spectra, all the
HOMO states correspond to nondegenerate $B_{1g}$ states. Also all $\alpha$-LUMO (lowest
unoccupied molecular orbital) states are the nondegenerate IR 11$A_{1g}$. It should be
noted that the $\alpha$-(HOMO-LUMO) gap is reduced considerably in DFT respect UHF, from
11.02 eV to 2.4 eV for the same basis. This result is consistent with the well known
fact that the HF method overestimates the HOMO-LUMO gap in many molecular systems.

\begin{figure}[th]
\centerline{\psfig{file=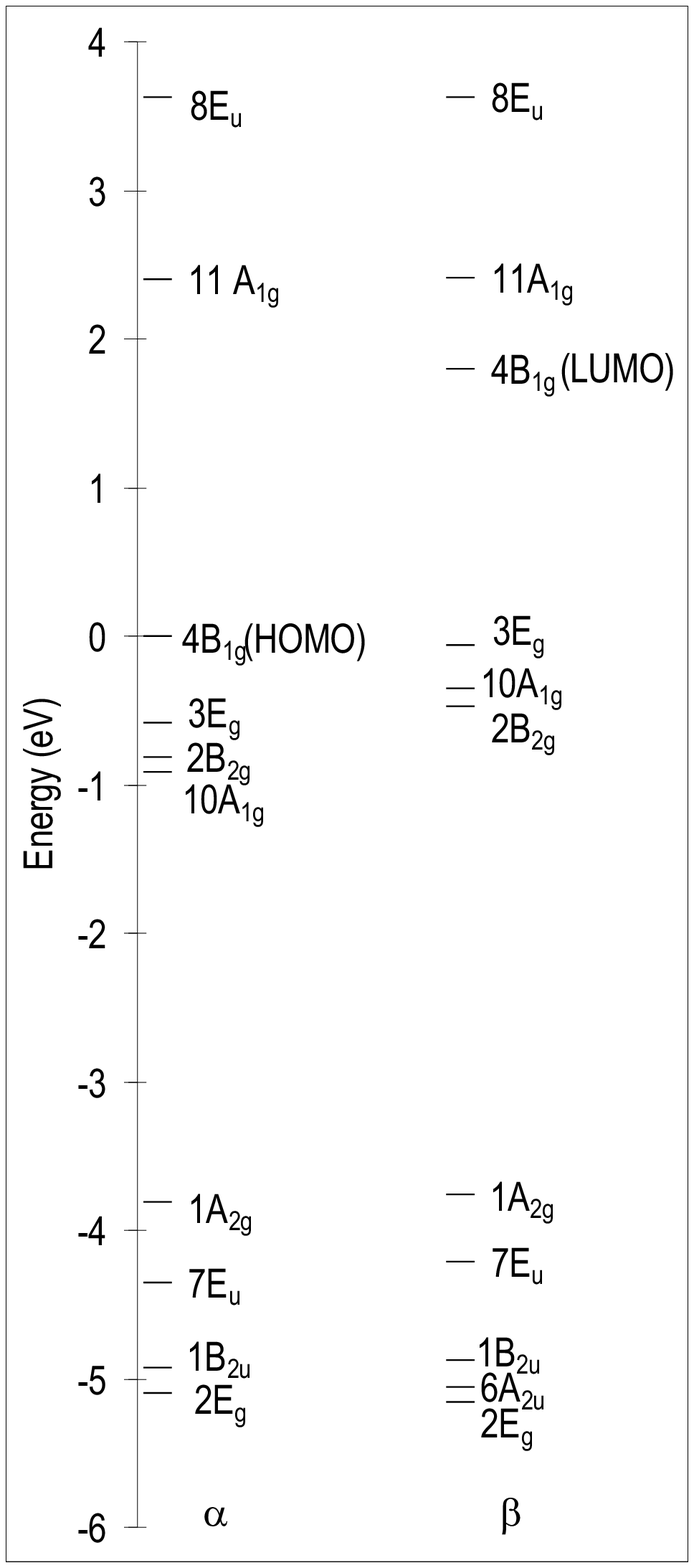,width=1.\textwidth}}
\vspace*{8pt}
\caption{Spin-unrestricted DFT-GGA spectrum for the CuO${_{6}}^{-10}$ cluster using the
Cu-3-21G plus O-4-21G$^{*}$ basis set.}
\label{F2}       
\end{figure}

\begin{table}[h]

\tbl{\label{tab:1} Square of the copper 3d and the oxygen 2p (planar and apical)
expansion coefficients of the highest occupied KS-BP86 MO's of the CuO${_{6}}^{-10}$
cluster.}
{\begin{tabular}{|c|c|c|c|c|c|c|c|c|c|c|c|} \hline
& d$_{xx}$ & d$_{zz}$ & d$_{xy}$ & d$_{xz}$ & d$_{yz}$ & p$_{x}$%
(p) & p$_{y}$(p) & p$_{z}$(p) & p$_{x}$(a) & p$_{y}$(a) & p$_{z}$(a) \\
\hline\hline $\alpha$ MO &  &  &  &  &  &  &  &  &  &  &  \\
\hline\hline 4B$_{1g}$ & 0.3546 & & & & & 0.1282 & 0.1282 & & &  &  \\ \hline
3E$_{g}$(1) &  &  & & 0.9354 & & & & 0.0621 & 0.0011 &  &  \\ \hline 3E$_{g}$(2) & &  &
&  & 0.9354 &  &  & 0.0621 &  & 0.0011 &  \\ \hline
2B$_{2g}$ &  &  & 0.9113 &  &  & 0.0432 & 0.0432 &  &  &  &  \\
\hline 10A$_{1g}$ & 0.1089 & 0.3313 &  &  &  & 0.0082 & 0.0082 & & &  & 0.0638 \\ \hline
1A$_{2u}$ &  &  &  &  &  & 0.5000 & 0.5000 & &  &  &  \\ \hline 7E$_{u}$(1) & &  &  &  &
& 0.9632 &  & & & &
\\ \hline 7E$_{u}$(2) &  &  &  & & & & 0.9632 & & &  &
\\ \hline 1B$_{2u}$ &  &  &  &  &  &  &  & 1.0000 & &  &
\\ \hline 2E$_{g}$(1) &  &  &  & 0.0267 &  &  & & 0.6081 &
0.3649 &  &  \\ \hline 2E$_{g}$(2) &  &  &  &  & 0.0267 &  & &
0.6081 &  & 0.3649 & \\ \hline\hline $\beta$ MO &  &  &  &  &  &  &  &  & & &  \\
\hline\hline 4B$_{1g}$ & 0.3971 & & & & & 0.0831 & 0.0831 & & &  &  \\ \hline
3E$_{g}$(1) &  & & & 0.9430 &  &  & & 0.0543 & 0.0012 &  &
\\ \hline 3E$_{g}$(2) &  &  &  &  & 0.9430 &  &  & 0.0543 &  &
0.0012 &  \\ \hline 10A$_{1g}$ & 0.1039 & 0.3056 &  &  & & 0.0100 & 0.0100 &  &  &  &
0.0564 \\ \hline 2B$_{2g}$ &  &  & 0.9166 & & & 0.0406 & 0.0406 &  &  &  &  \\ \hline
1A$_{2u}$ & &  &  & & & 0.5000 & 0.5000 &  &  &  &  \\ \hline 7E$_{u}$(1) & &  & &  & &
0.9692 &  &  & 0.0008 &  &  \\ \hline
7E$_{u}$(2) &  &  &  & &  &  & 0.9692 &  &  & 0.0008 &  \\
\hline 1B$_{2u}$ &  & &  &  &  &  &  & 1.0000 &  &  &  \\
\hline 2E$_{g}$(1) &  &  &  & 0.0226 &  &  &  & 0.6342 & 0.3429 & &  \\ \hline
2E$_{g}$(2) &  &  &  &  & 0.0226 &  &  & 0.6342 & & 0.3429 &  \\ \hline
\end{tabular}}
\end{table}

In Fig.~\ref{F2} we show the DFT spectra for $\alpha$ and $\beta$ states and in
Table~\ref{tab:1} the contributions from the copper 3d and the planar and apical oxygens
2p to the highest occupied KS-BP86 MO's. The LUMO is the $\beta$ state 4$B_{1g}$ located
1.8 eV over the HOMO, compared with 1.4 eV reported in Ref.~\refcite{Husser}. We would
like to remark some of the particular features shown in the DFT spectrum. There are two
molecular orbital ``bands'' in the occupied electronic spectrum under the HOMO,
separated by a 2.9 eV gap. The first band with energy between 0 and -0.91 eV have MO's
with predominantly 3d cooper components: 4B$_{1g}$ (HOMO), 3E$_g$, 2B$_{2g}$, and
10A$_{1g}$. The second band with energy between -3.81 and -5.10 eV include MO's composed
mainly by oxygen 2p components: 1A$_{2g}$, 7E$_u$, 1B$_{2u}$, and 2E$_g$
(Table~\ref{tab:1}). This feature is not present in the UHF spectrum, where this MO
bands are mixed. Similar results were obtained in Ref.~\refcite{Husser}: the first band
between 0 and -1.5 eV and the second band between -2.6 and -4.4 eV.

\begin{figure*}[th]
\centerline{\psfig{file=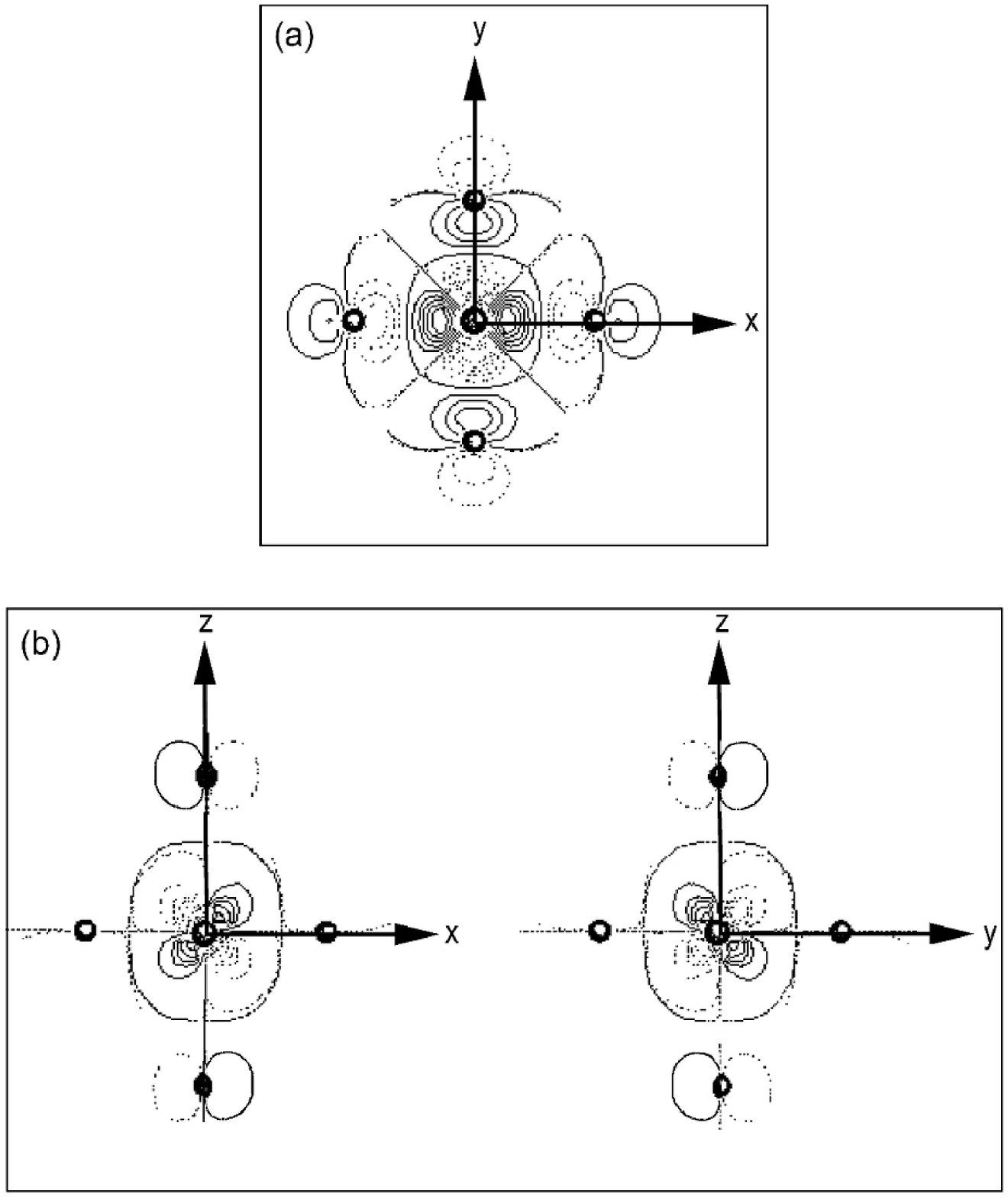,width=.5\textwidth}}
\centerline{\psfig{file=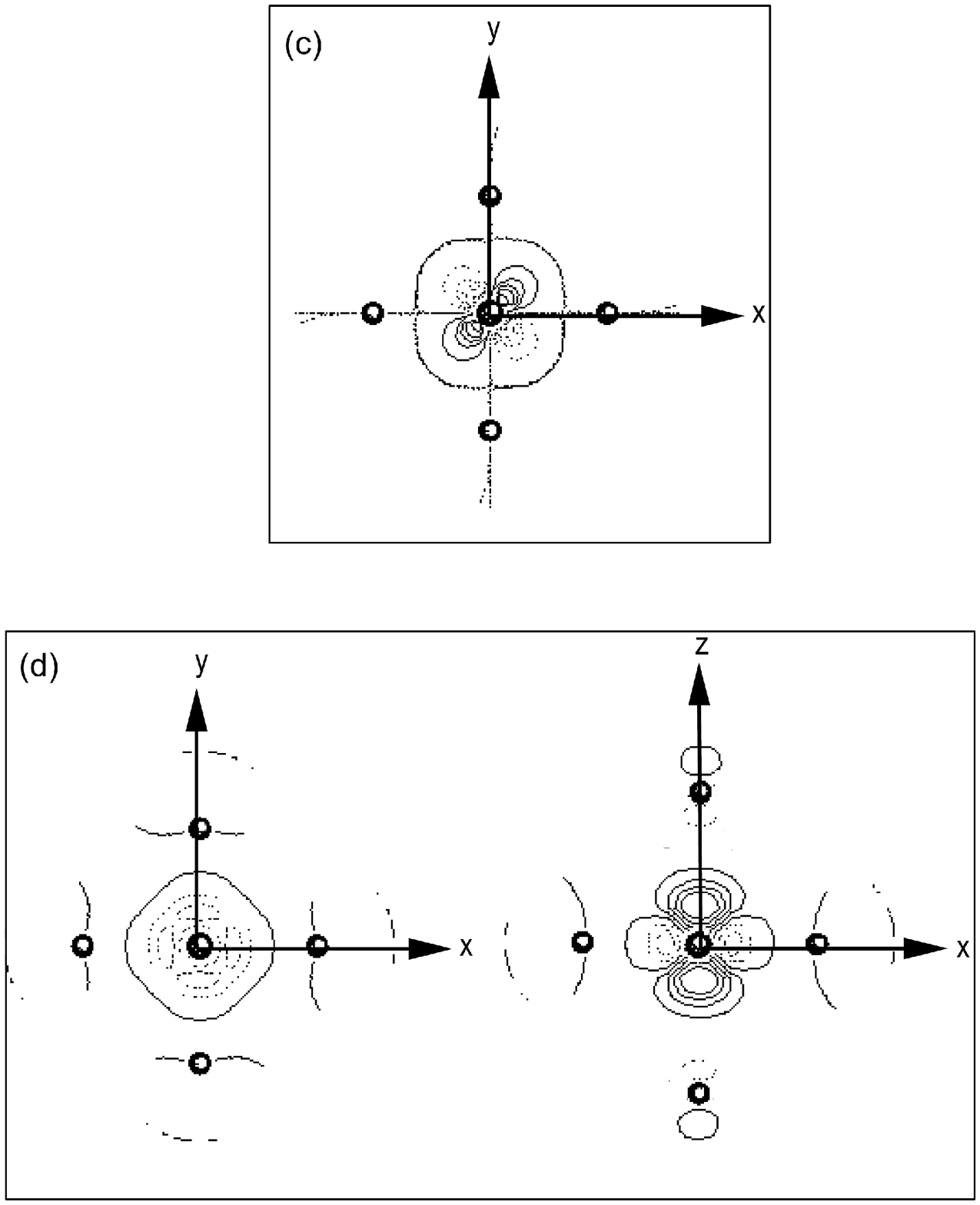,width=.5\textwidth}}
\vspace*{8pt}
 \caption{Hybridzation of the DFT-GGA wave function
for the predominantly d orbitals of the CuO${_{6}}^{-10}$ cluster: (a) 4$B_{1g}$ (HOMO),
(b) 3$E_g$, (c) 2$B_{2g}$, and (d) 10$A_{1g}$.} \label{F3(a-d)}
%
\end{figure*}

 We show in Fig. \ref{F3(a-d)} the predominant d orbitals for the 4B$_{1g}$ (HOMO), 3E$_g$,
2B$_{2g}$, and 10A$_{1g}$ containing the 3d$^{9}$ states of Cu$^{2+}$. In the HOMO state
4B$_{1g}$, the 3d$_{x^2-y^2}$ state of Cu is hybridized with the 2p$\sigma$ O orbitals
along the Cu-O bond in the x-y plane. This state is occupied only by one $\alpha$
electron and is orbitally nondegenerate. It should be noted however that it is spin
degenerate, but since it has a degeneracy of order two in spin, it does not produce a JT
deformation.\cite{J} It is basically the same kind of hybridization we got for the UHF
wave function, except by a lower contribution from the 2p$\sigma$ O orbitals. The
occupied state nearest to the HOMO, the double degenerate orbital 3E$_g$, could be seen
as the result of hybridization of the 3d$_{xz}$ (3d$_{yz}$) states with the p$_x$ and
$-$p$_x$ (p$_y$ and $-$p$_y$) oxygen orbitals along the apical axis with no contribution
of the p$_z$ and -p$_z$ oxygen orbitals along the x (y) axis. The 2B$_{2g}$ orbital
becomes the hybridization of the Cu 3d$_{xy}$ state with the p$\pi$ orbitals of the
oxygens in the x-y plane (perpendicular to the Cu-O bond). Finally, the 10A$_{1g}$
orbital is the hybridization of the 3d$_{z^2-r^2}$ state of Cu with the 2p$\sigma$
states of the oxygens in the apical z axis and with the p$_x$, $-$p$_x$, p$_y$ and
$-$p$_y$ orbitals of the oxygens in the x-y plane along the Cu-O bonds.

Hence, since the electronic ground state is the nondegenerate $^{2}$B$_{1g}$ state, we
can not have  a residual static JT effect in the CuO${_{6}}^{-10}$ cluster. The
possibility for a PJT deformation, in this state of oxidation, is low since the
HOMO-LUMO gap which is about 2 eV   is much greater than the characteristic vibrational
energy (20-100 meV). However, in the oxidized CuO${_{6}}^{-9}$ cluster, the spin-singlet
$^{1}$A$_{1g}$ electronic ground state could be mixed with some near excited states
giving non-zero PJT vibronic couplings. Those states could be the spin-singlet
(predominantly copper 3d states) $^{1}$E$_{g}$ and/or $^{1}$B$_{1g}$. This result is
consistent with the assumption made in the model introduced by Bersuker\cite{Bersuker2}
to explain a polaron formation mechanism in high T$_{c}$ superconductors. There he
assumes a PJT vibronic coupling through the mixing with the $^{1}$B$_{1g}$ excited
state. Here we have shown that besides the $^{1}$B$_{1g}$ excited state, the inclusion
of the $^{1}$E$_{g}$ excited state in a PJT vibronic coupling might be also relevant. In
the context of the model we have presented in this work, the PJT effect may produce a
coupling with the e$_{g}$ and/or b$_{1g}$ vibrational modes which could be responsible
for the local deformations in the superconducting system.

Recent DFT Becke-3-Lee-Yang-Parr band calculations\cite{Perry} in the
La$_{2-x}$Sr$_{x}$CuO$_{4}$ doped system support a highly inhomogeneous hole formation
with symmetry $^{1}$A$_{1g}$ adjacent to the Sr impurity, giving us additional
reliability about our final conclusion.

\section{Conclusions}

\label{conclusions}

We have performed {\it ab initio} calculations in the{\it \ }CuO${_{6}}^{-10}$ elongated
octahedra cluster having the symmetry group D$_{4h}$, representing the La-Ba-Cu-O
system, to investigate the possibility of having a Jahn-Teller or pseudo Jahn-Teller
effect as the origin for the local distortions found in this system. This study has been
motivated by the fact that the apical distortion of the CuO$_{6}$ octahedra in the
insulating parent compound of the La based superconductor La$_{2}$CuO$_{4}$ is only
partially due to a static Jahn-Teller effect, having therefore a {non {\it a priori}
condition to remove all the degeneracy of the electronic ground state of the parent
regular octahedron with O$_{h}$ symmetry.}

The DFT-GGA electronic spectrum was compared with the corresponding UHF calculated in a
previous work. As a result in both cases, we found that the electronic degenerate states
correspond only to intermediate levels, which are fully occupied. The symmetry of the
total electronic ground state wave functions being therefore a nondegenerate
$^{2}$B$_{1g}$ state, ruling out the possibility of a residual static JT effect in the
CuO${_{6}}^{-10}$ cluster. It should be noted however, that the DFT electronic spectrum
contains a degenerate 3E$_{g}$ state very close to the HOMO 4B$_{1g}$ state. This
belongs to a first set of filled molecular levels, including the HOMO state, which are
spread on a width of the order of 1 eV, and which are characterized by being
predominately formed by the 3d states of Cu. This set is separated by a gap of the order
of 3 eV from another set of molecular states originated mainly by 2p states of O.

Since the HOMO-LUMO gap (about 2 eV) is much greater than the characteristic vibrational
energy (20-100 meV), the possibility for a PJT deformation, in this state of oxidation
is low. However, in the oxidized CuO${_{6}}^{-9}$ cluster, the spin-singlet
$^{1}$A$_{1g}$ electronic ground state could be mixed with some near excited states
giving non-zero PJT vibronic couplings. Those states could be the spin-singlet
(predominantly copper 3d states) $^{1}$E$_{g}$ and/or $^{1}$B$_{1g}$. We therefore
conclude that the PJT effect in  the CuO${_{6}}^{-9}$ cluster may produce a coupling
with the e$_{g}$ and/or b$_{1g}$ vibrational modes. This PJT effect could be the
responsible for the local distortions observed in the CuO$_{6}$ octahedra in the
La$_{1.85}$Sr$_{0.15}$CuO$_{4}$ superconductor.

\end{document}